\documentclass[]{aastex631}

\usepackage{color}
\usepackage{amssymb}

\shorttitle{Small-Scale Interchange Reconnection}
\shortauthors{Mason \& Uritsky}

\graphicspath{{./}{figures/}}

\begin{document}

\title{Statistical Evidence for Small-Scale Interchange Reconnection at a Coronal Hole Boundary}


\correspondingauthor{Emily I. Mason}
\email{emason@predsci.com}

\author[0000-0002-8767-7182]{Emily I. Mason}
\affiliation{Predictive Science Inc. \\
9990 Mesa Rim Rd, Suite 170 \\
San Diego, CA 92121, USA}

\author[0000-0002-5871-6605]{Vadim M. Uritsky}
\affiliation{Catholic University of America \\
620 Michigan Ave., N.E. \\
Washington, DC 20064, USA}
\affiliation{NASA Goddard Space Flight Center \\ 8800 Greenbelt Ave.\\Greenbelt, MD 20771, USA}

\begin{abstract}

Much of coronal hole (CH) research is focused upon determining the boundary and calculating the open flux as accurately as possible. However, the observed boundary itself is worthy of investigation, and holds important clues to the physics transpiring at the interface between the open and closed fields. This Letter reports a powerful new method, an application of the correlation integral which we call correlation dimension mapping (CDM), by which the irregularity of a CH boundary can be objectively quantified. This method highlights the most important spatial scales involved in boundary dynamics, and also allows for easy temporal analysis of the boundary. We apply this method to an equatorial CH bounded on two sides by helmet streamers and on the third by a small pseudostreamer, which we observed at maximum cadence for an hour on 2015 June 4. We argue that the relevant spatial scales are in the range of $\sim 5-20$ Mm, and we find that boundary complexity depends measurably upon the nature of the neighboring closed structure. The boundary along the pseudostreamer shows signs of highly-localized, intermittent complexity variability, likely associated with abrupt changes in the magnetic topology, which would be elegantly explained by interchange reconnection. By contrast, the helmet streamer boundary supports long-lived high-complexity regions. These findings support the recent predictions of interchange reconnection occurring at very small scales in the corona. 

\end{abstract}

\section{Introduction} \label{sec:intro}

Since the early 20th century, coronal holes (CH) have been defined observationally as areas of low emission off-limb in white light during eclipses \citep{Cranmer2009}. The definition was extended to areas of low emission in the X-ray and ultraviolet regimes at the start of the Space Age, when images of the Sun showed large dark patchy regions on-disk \citep{Munro1972,Altschuler1972}. The association of these observational signatures with magnetic open fields came shortly afterwards, as initially discussed by \citet{Pneuman1973}.

Physicists soon discovered a disconnect when comparing the amount of magnetic flux bounded by these dark regions to the measurements taken by in-situ instruments. Now known as the ``open flux problem", many studies have found that in-situ measurements lead to estimates of open flux in the heliosphere that are several times higher than those derived from magnetograms and CH maps \citep{Linker2017,Riley2019,Wang2022}. It is currently unknown if this is a problem with magnetographs themselves, as magnetic field readings are frequently non-uniform across various instruments \citep{Riley2014,Wallace2019,Wang2022}. It could also be due to the highly complex structure of the low corona, coupled with observational limitations along a single line of sight, which both serve to obscure the physical open/closed boundary \citep{Kirk2009,Linker2021,Reiss2021}, or even to misinterpretations of the in-situ open flux measurements \citep{Frost2022}. This problem has driven an active research area devoted to determining CH boundaries as best we can using remote data \citep{Aschwanden2005, Esser1999,Krista2011,Garton2018,Reiss2021}. There are practical motivations to constrain a definition that is both observationally detectable and physically motivated: the high-speed wind streams from CHs are among the most important drivers of space weather, and forecasters are routinely searching for ways to make arrival times more accurate.

One recent study posited, through a 3D MHD simulation, that random driving at helmet streamer and pseudostreamer CH boundaries would produce distinct, persistent, and observable differences in the boundary \citep{Aslanyan2022}. The argument for this is that interchange reconnection (initiated by random motions at the supergranule scale) occurs more readily at pseudostreamer boundaries. There, the magnetic stresses are processed through the relatively low-lying x-point, with the result that the boundary is rapidly smoothed after reconnection. By contrast, more magnetic stresses can accumulate along the helmet streamer boundary due to the longer field lines. They found the time expected for these differences to become apparent to be several days. In observations, a similar explanation has been put forward by in-situ analysis of events that may have originated at open/closed boundaries \citep{Crooker2014,Owens2018,Macneil2020}. The CH we study here is topologically similar to the structure modeled in that paper, and our findings support and expand upon their theoretical conjectures, as will be discussed in greater detail in Section \ref{sec:disc}.

Despite the major efforts being dedicated to CH research, very little work has been directed at quantifying or analyzing the observable boundaries themselves. It is apparent when looking at observations that some CH boundaries are much more irregular than others; sometimes even the various sides of the same CH appear quite different (see Figure \ref{F1}a). Projection effects do not entirely account for this difference, as the appearance can persist before and after crossing disk center. This leaves an underlying physical cause. What, then, drives this irregularity, and how can we quantify it? Can it inform our investigations into the physics of CH boundaries, and what might the effects of these differences be on the solar wind? 

These are the driving questions that motivate this Letter. By introducing a new and versatile image processing method -- the \textit{correlation dimension mapping} (CDM) -- we quantify, for the first time, spatially localized irregularities in a CH boundary, and measure the most relevant physical scales responsible for short-term changes in the boundary. Our findings demonstrate that a critically important aspect of the CH morphology associated with abrupt magnetic field reconfigurations can be subject to rigorous quantitative analysis, allowing for deeper understanding of the drivers and dynamics at the interface regions of the open and closed corona.

\section{Data}\label{sec:data}

\begin{figure}
\centering
    \includegraphics[width=1.\linewidth,clip]{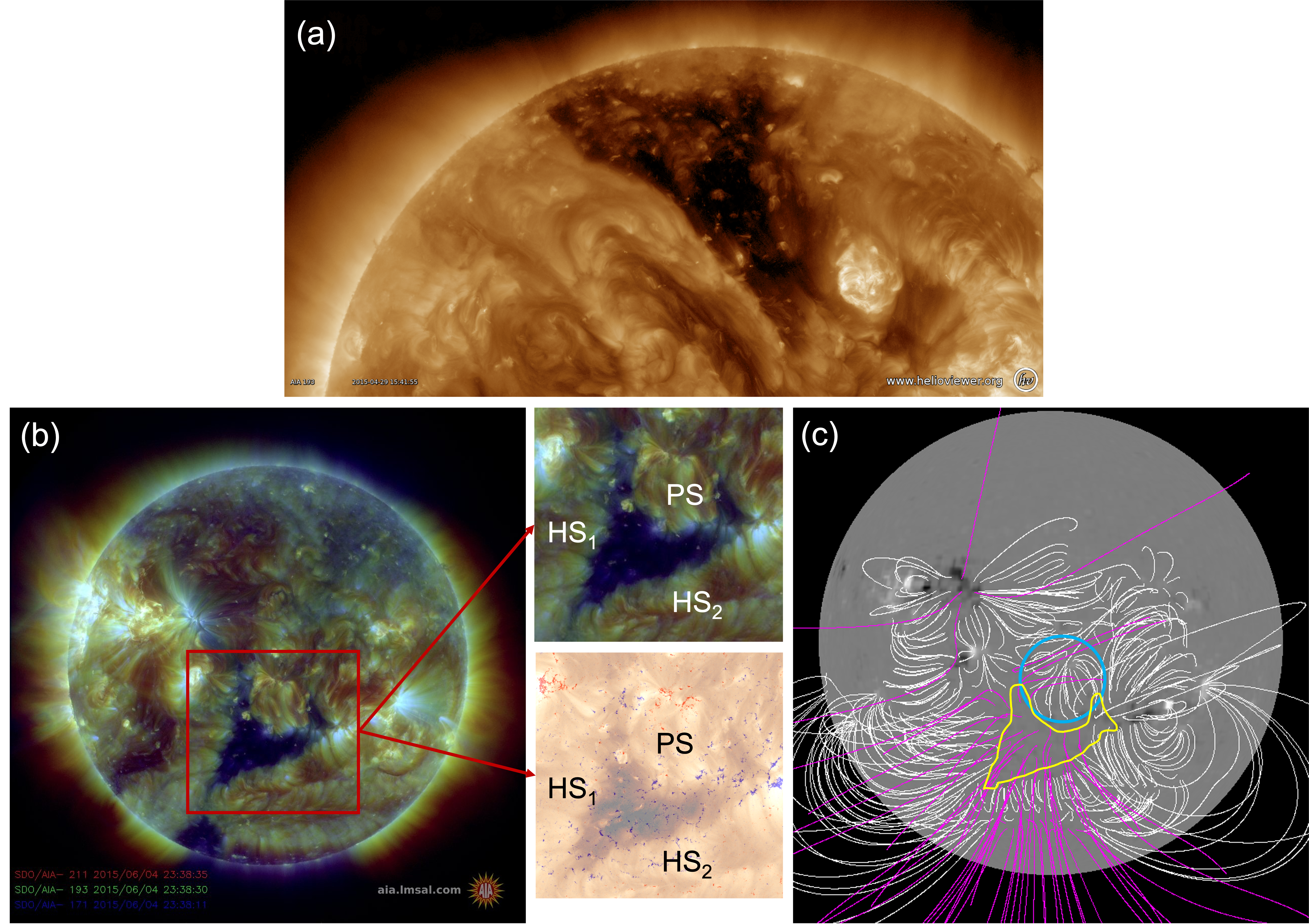}
              \caption{(a) SDO AIA 193 Å image of a northern polar CH; note the much more pronounced and smoother appearance of the boundary on the western side, compared to the less-distinct and patchier boundary on the eastern side. (b) SDO composite 171/193/211 Å context image showing the full Sun on 2015 June 4, with a red box showing the neighborhood of the insets. The top inset shows the CH with the three closed structures bounding it, namely two helmet streamer edges and one pseudostreamer edge. The bottom inset shows the 193 Å data paired with a color-coded HMI magnetogram. (c) PFSS model from the Solarsoft tool pfss$\_$viewer, estimating the extrapolated open and closed field on the same date; the yellow outline is the approximate boundary of the observed CH, and the blue circle highlights the anemone-like topology of the pseudostreamer structure on the hole's northern border.}
              \label{F1}
\end{figure}

The CH we selected for this investigation was near disk-center on 2015 June 4 (see Figure \ref{F1}b). We used Solar Dynamics Observatory Atmospheric Imaging Assembly \citep[SDO AIA;][]{Lemen2011,Pesnell2012} 193 Å data, which is dominated by plasma around 1.5 MK. We conducted standard processing to bring the data up to level 1.5 using the SunPy package \citep{sunpy_community2020}, and reviewed 153 images at 24-second cadence for the hour from 22:30 to 23:30 UTC. No large-scale events occurred near the CH during this time frame (e.g., large flares, coronal mass ejections, large jet eruptions, etc.). One small jet eruption occurs within the coronal hole, which is covered in the analysis since it affects the boundary.

Our criteria were to select an equatorial CH with the simplest bordering closed structures, that passed as close to disk center as possible. This ensures that the projection effects were minimized as much as possible. We also looked for a CH that was not immediately adjacent to a large active region or prominence; either of these could affect the boundary region itself, or host complicating short-term dynamics like eruptions. This CH met all of those requirements.

The structures of the bounding regions were determined by inspection with several data sets and extrapolations; the SDO HMI data \citep{Scherrer2012} showed that the structure on the northern border was indeed formed around an isolated and localized minority polarity, providing a large ``anemone jet" \citep{Shibata1994} or, equivalently, a small pseudostreamer. The  potential field source-surface (PFSS) model seen in Figure \ref{F1}c indicate that the southern and western boundaries of the CH are connected to helmet streamers.

\section{Correlation Dimension Mapping}\label{sec:CDM}

\begin{figure}
\centering
    \includegraphics[width=1.\linewidth,clip]{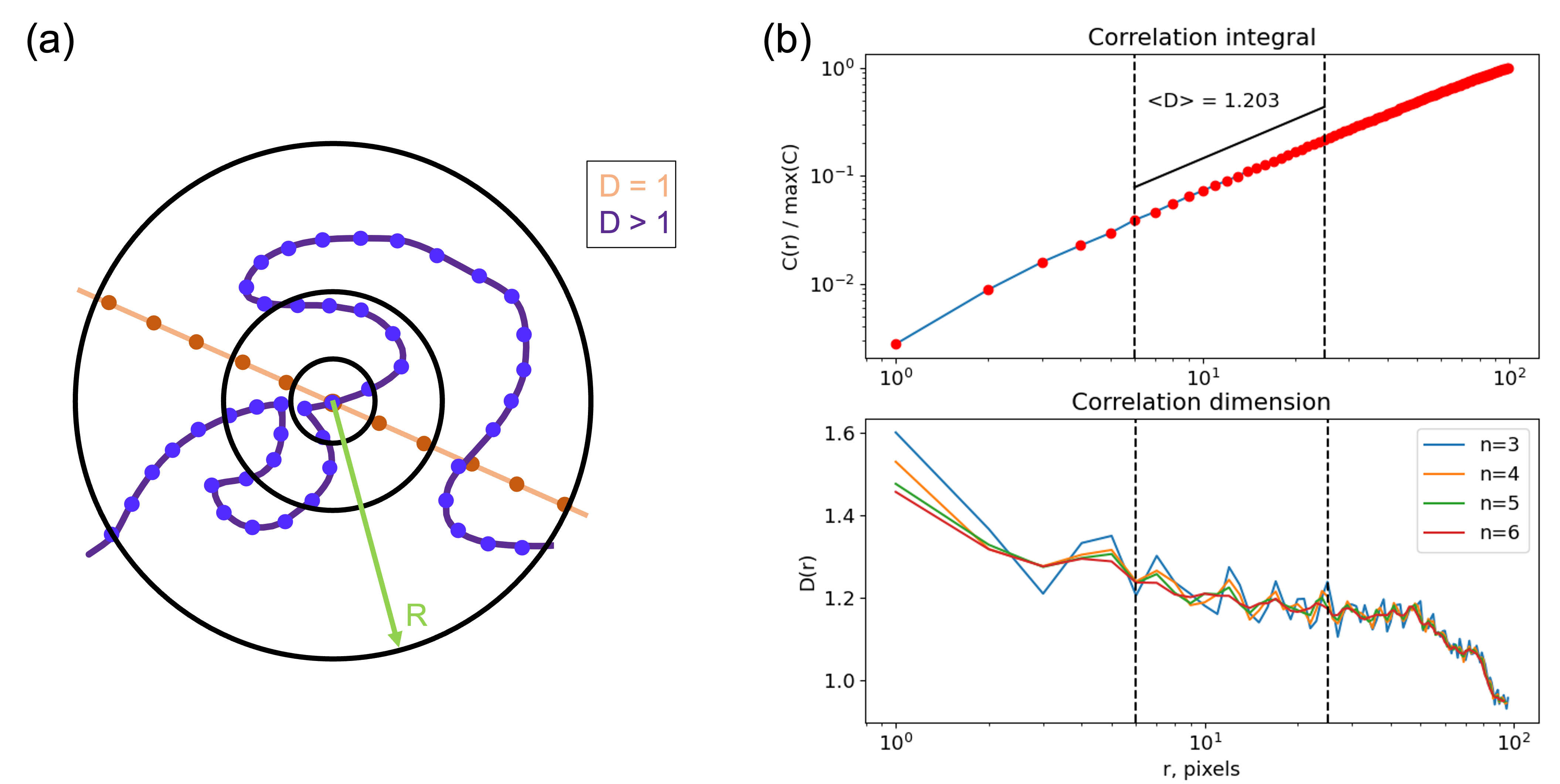}
              \caption{(a) Diagram illustrating the basic concept of the correlation integral; concentric circles of increasing radius encompass a linearly-increasing number of points on the line (points shown in orange), but a greater-than-linearly increasing number of points on the purple line (points in blue). (b) Top: the log-log plot of the correlation integral (the normalized number of points encompassed in the CH boundary, averaged over all processed image frames) vs. the bounding circle radius; the dotted black lines between 6 and 25 pixels show the region with the most relevant physical scales for this boundary. Bottom: log-linear plot of the scale-dependent correlation dimension obtained using a sliding window of variable size $n$ vs. the bounding circle radius; the same scales are highlighted as in the above plot for clarity.}
              \label{F2}
\end{figure}

Many solar structures have complex morphologies which could be studied using a variety of available data analysis methods \citep[e.g.,][and references therein]{Georgoulis12, Vlahos16, Aschwanden22}. To obtain a spatially-dependent measure of irregularity of the CH boundary, we introduce a new statistical technique which generalizes the method of the correlation integral introduced by \citet{Grassberger1983}. The method was initially designed as a proxy to the Hausdorff dimension of a random set of points embedded in a space of dimension $d$, such as a set of discrete phase space coordinates representing an attractor of a nonlinear dynamical system \citep{Schuster05}. The correlation integral characterizes the probability of finding a pair of data points within a hypersphere of a specified radius $r$, and is approximated by the sum
\begin{equation}
C(r) = \sum_{j=i-1}^N \sum_{i=1}^N \Theta (r - \|\mathbf{q_i} - \mathbf{q_j}\|),
\label{eq:CI}
\end{equation}
where $\mathbf{q_i}$ and $\mathbf{q_j}$ are the $d$-dimensional coordinates of the data points indexed with $i$ and $j$, $N$ is the number of points in the set, $\Theta$ is the Heaviside step function, and the summation is performed over all distinct pairs of points belonging to the studied set. For a self-similar set of points, the $C(r)$ dependence takes the power law form
\begin{equation}
C \sim r^D,
\label{eq:D}
\end{equation}
in which $D \leq d$ is the correlation dimension \citep{Grassberger1983} providing a quantitative measure of the set's geometry. Integer $D$ values are obtained when the points are arranged into regular $D$-dimensional shapes such as e.g. smooth lines ($D=1$), surfaces ($D=2$), or volumes ($D=3$), with the upper limit imposed by the embedding dimension $d$. Fractional $D$ values signal a self-similar clustering of the data points across the studied range of $r$ scales. In general, the higher the fractional $D$, the more complex is the geometry of the clustered data set under investigation \citep{Uritsky12, Uritsky14}. This is illustrated by Figure \ref{F2}a; this shows the difference between a completely straight line and an arbitrarily curved one.

While the correlation dimension has important theoretical implications for dynamical system analysis (see e.g. \citet{Schuster05} and references therein), in this study we invoke this concept with a simple practical purpose in mind -- as an objective and statistically robust method for quantifying the irregularity of the CH boundary. Since the complexity of different portions of the boundary line can vary significantly depending on the local magnetic geometry, we introduce a spatially-dependent version of the correlation integral 
\begin{equation}
C(r,x,y) = \sum_{i=1}^{N(x,y)} \Theta \left(r - \sqrt{(x_i-x)^2 - (y_i-y)^2}\right),
\label{eq:CI_loc}
\end{equation}
where  $\{ x_i, y_i \} \equiv \{ \mathbf{q_i}\} $ are the rectangular coordinates of the CH boundary extracted from a two-dimensional coronal image ($d=2$). In contrast to the original correlation integral definition (\ref{eq:CI}) that represents a counting statistics of {\it all} pairs of points lying within a specified distance $r$ from each other, our modified definition (\ref{eq:CI_loc}) operates on a {\it subset} of these points surrounding a given reference location $(x,y)$. The radius of the studied neighborhood is defined by the largest scale $r_{max}$ used in the calculation, and the total number $N$ of points included in the sum depends on the location. A restriction is applied that confines the points considered to those within a set maximum distance along the boundary, to exclude regions that are spatially close but distant within the boundary array, and so are topologically unrelated. This is illustrated by the white box of Figure \ref{F4}d, where a yellow and blue segment nearly touch but do not affect each other.

Once the data cube $C(r,x,y)$ is computed for a given boundary line, the spatially-dependent correlation dimension is calculated by approximating (\ref{eq:CI_loc}) with a locally defined power law: 
\begin{equation}
C(r,x,y) \sim r^{D(x,y)}.
\label{eq:D_loc}
\end{equation}

Technically, the local power-law slope $D(x,y)$ is evaluated by fitting a straight line to the $C(r)$ dependence plotted in double logarithmic coordinates across a finite range of scales $r \in [r_1, r_2]$ selected based on the goals of the analysis, within the limits imposed by the image resolution and the size $r_{max} \geq r_2$ of the local neighborhood. The resulting $D(x,y)$ array is the CDM representing position-dependent irregularity of the studied boundary line.

\section{Results}\label{sec:results}

\begin{figure}
\centering
    \includegraphics[width=.8\linewidth,clip]{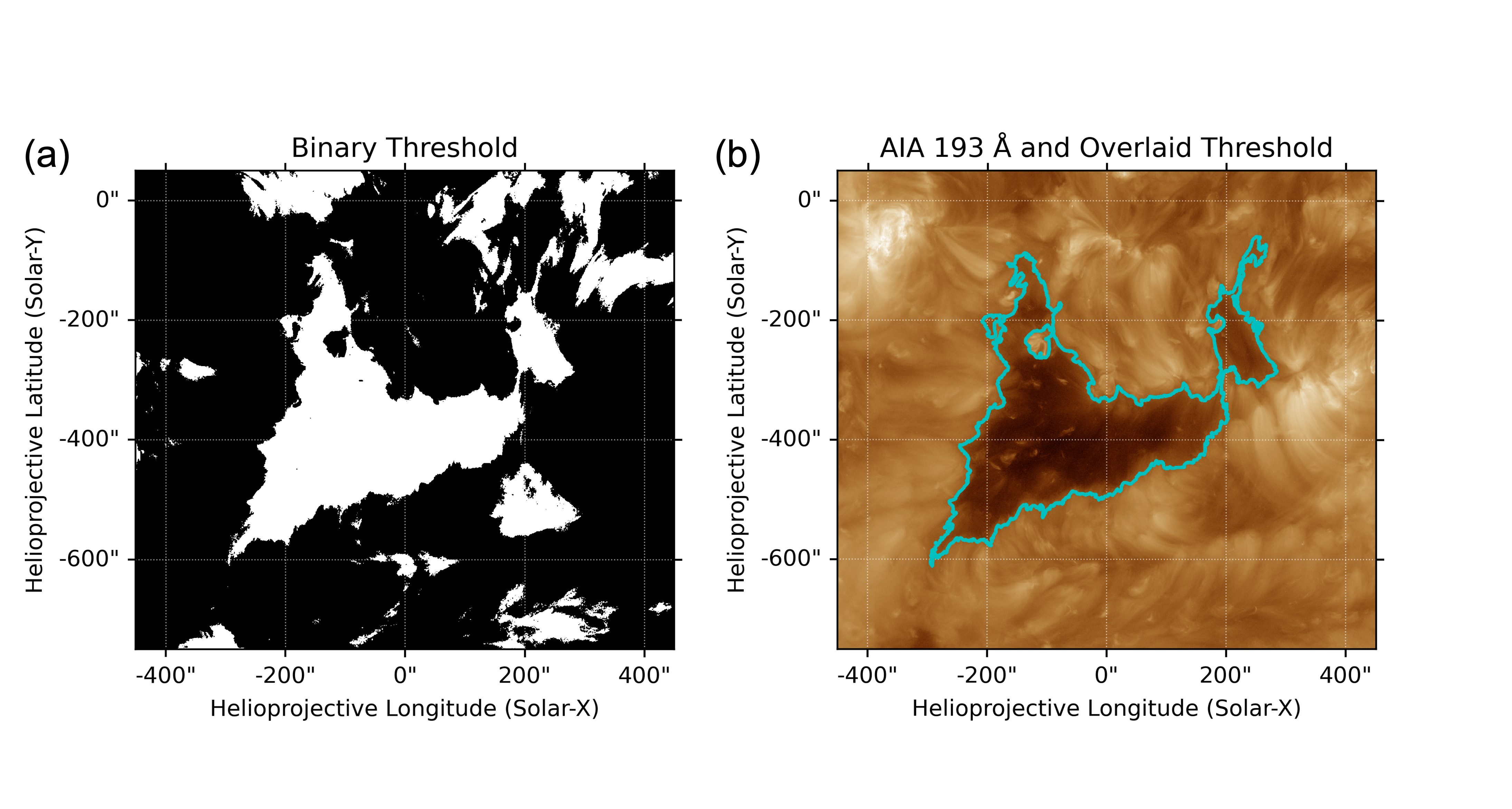}
              \caption{(a) Binary threshold image of the cutout taken at value $log(DN)=2$. (b) Extracted boundary of the CH overlaid on the cutout in orange.}
              \label{F3}
\end{figure}

To prepare the observations for assessment with the CDM method, we used an 8-point median smoothing method on the cutouts and constructed binary images at a threshold value of 100 DN (Figure \ref{F3}a). This allowed the contouring procedure to easily pick out the largest shape as the boundary of the CH (Figure \ref{F3}b). 

We selected the 6-25 pixel range of scales for the mapping via both statistical and physical arguments. Statistically, the correlation dimension graphs of the studied CH images (Figure \ref{F2}b) exhibit a region of r values with an approximately constant D. We selected the 6-25 pixel subrange, which equals 5-21 Mm (each AIA pixel is approximately 435 km in width). The minimum size here aligns very well with the characteristic sizes of plumelets, believed to be low-lying hosts of high-frequency interchange reconnection, and supergranules at the large end, which are frequently credited with generating magnetic reconfiguration at the edges where flows converge. We discuss this further in Section \ref{sec:disc}.

\begin{figure}
\centering
    \includegraphics[width=1.\linewidth,clip]{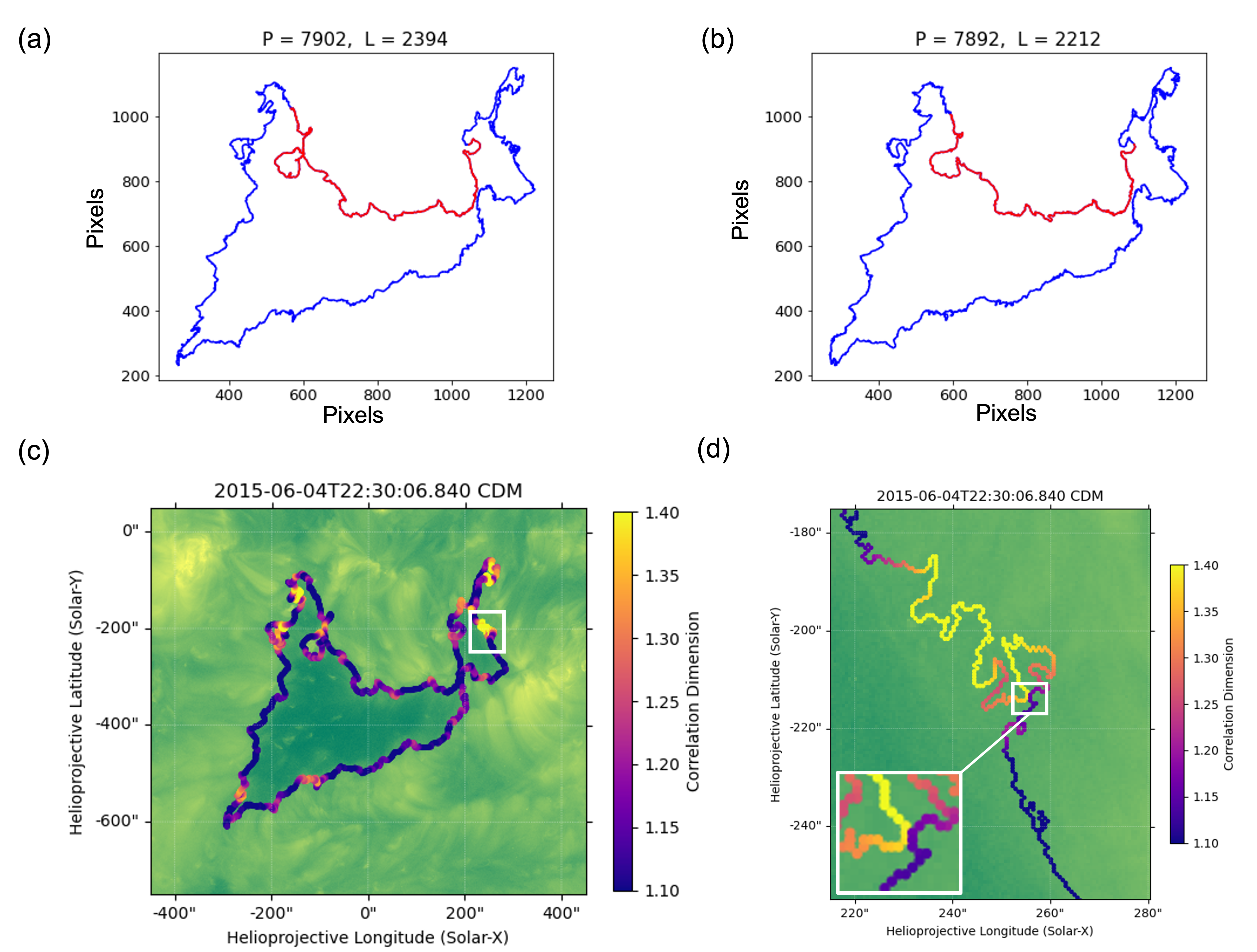}
              \caption{(a) Perimeter map of threshold boundary at 22:30 on 2015 June 4, showing P (the full perimeter in pixels) and L (the length of the region along the pseudostreamer boundary). (b) The same map, at 23:30. (c) Correlation dimension map overlaid on the initial cutout; blue  denotes straighter boundaries, while yellow corresponds to more irregular boundaries. The inset shown in (d) is outlined in white. (d) An enlargement of the CDM shown in (c); at a greater level of detail it is easier to see the small-scale variations in the boundary.}
              \label{F4}
\end{figure}

Figure \ref{F4}a and \ref{F4}b show the boundary extraction and mapping procedures; the perimeter of the coronal hole with the pseudostreamer portion of the boundary is highlighted in red at the first and last time step analyzed (22:30 and 23:30). The full perimeter length is 7902 and 7892 pixels, respectively, a reduction of less than 2\%. The pseudostreamer portion of the boundary also shrank in length, going from 2394 to 2212 pixels (7\% decrease).  Figure \ref{F4}c exhibits the CDM procedure applied to the first cutout; the color map shows higher dimension values closer to the yellow end of the spectrum, while lower dimension values are closer to purple. The box highlights a portion of the boundary which illustrates the utility of CDM. The color map has been slightly saturated to show the trends more clearly; the actual range of values over the entire boundary was 0.85 to 1.59.

It is evident that the CDM plot has a surprisingly high level of detail (Figure \ref{F4}d), the enlarged portion of the previous panel. This region makes it easier to interpret that straighter portions of the boundary are dark purple (corresponding to lower dimension values), while those with that are more curved or bent are yellow (corresponding to higher dimension values). One of the strengths of the CDM method is that portions of the boundary which loop around for some distance and then return close to a different part of the line are not taken into account when computing the correlation dimension. This removes the ambiguity that would otherwise arise in an area like that in the easternmost portion of the CH boundary, where two bends of the line nearly meet and the bounding circles would have intersected both sides of the loop.

\begin{figure}
\centering
    \includegraphics[width=.7\linewidth,clip]{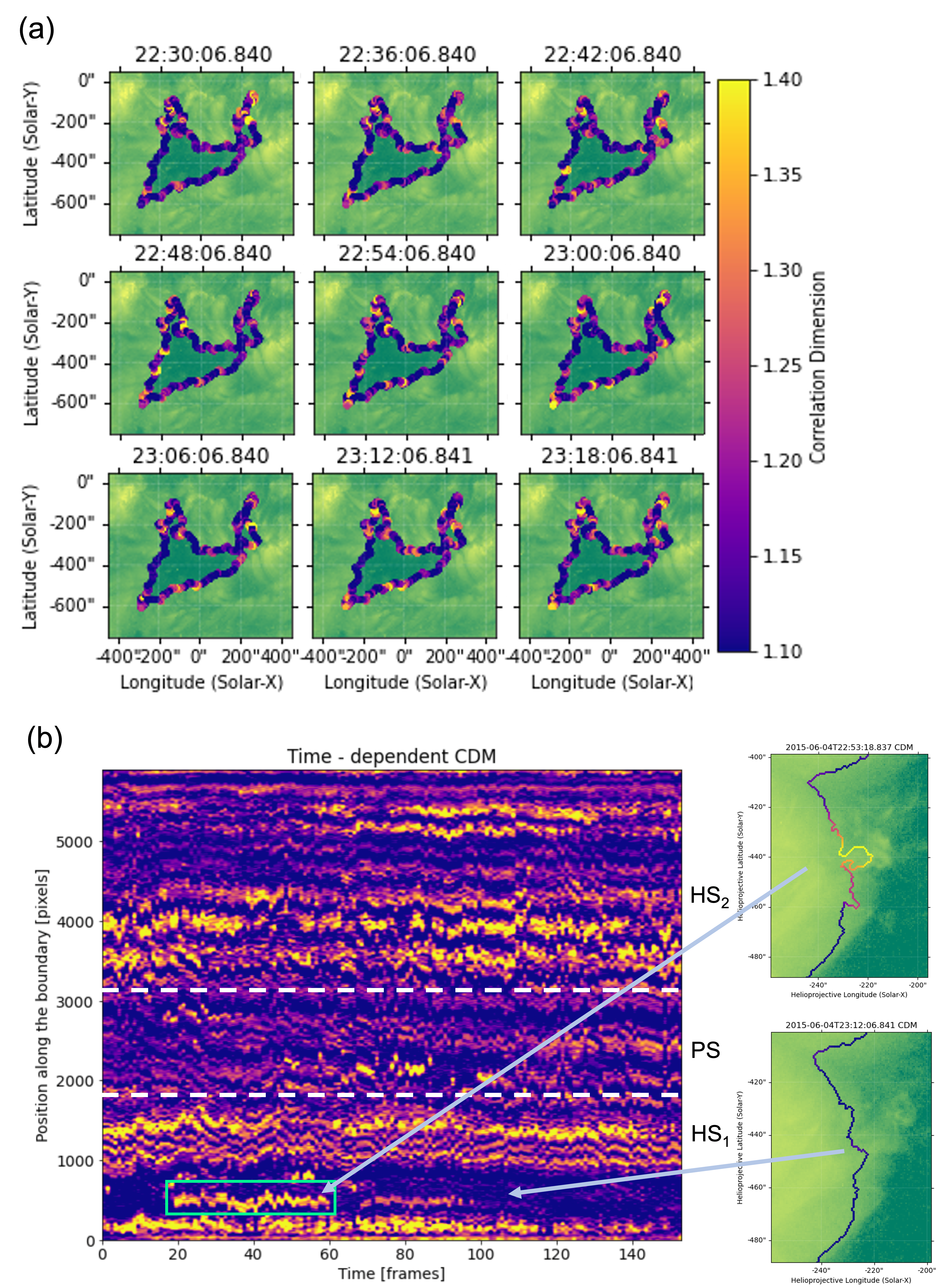}
              \caption{(a) Nine CH observations with their respective CDM overlays, spaced evenly across the hour of the investigation. The 193 Å data is mapped in green to enhance contrast with the overlay. A 6-second movie, available online, shows the full time evolution of the correlation dimension. (b) Time-distance plot showing the distribution of the correlation dimension along the boundary as a function of the image frame, normalized in length as explained in the text. Each vertical slice represents a CDM such as those shown in panel (a), with the same color scale. The white dotted lines designate the approximate pseudostreamer/helmet streamer boundaries, and the box highlights the boundary jet signature. The arrows indicate the temporal location of the insets, which show the evolution of the boundary during the jet eruption.}
              \label{F5}
\end{figure}

We applied the CDM analysis to all 153 observations in the selected hour-long span. The results of a few of these mappings are presented in Figure \ref{F5}a (the full set of maps are available online as a movie). It is clear from these maps that the regions of high complexity are spatially localized, and that they change from frame to frame. However, to understand how these changes temporally requires more thorough analysis, shown in panel \ref{F5}b. There is also no clear overall difference in the complexity between the boundaries adjoining the helmet streamers and the boundary adjoining the pseudostreamer; however, Figure \ref{F5}b tells a different story. This panel shows a length-normalized time-distance plot for the CH boundary over the hour of observation. The position along the boundary on this plot was measured in the clockwise direction, starting from the lower bottom corner of the boundary. The length of the boundary arrays obtained at different time steps was equalized using a linear interpolation implemented in the  {\tt interp1d} method of the {\tt scipy} Python package.

In contrast to the unprocessed boundary line, the CDM makes it obvious that the portion of the boundary marked out as adjoining the pseudostreamer (\emph{PS}) presents an overall smoother structure, and the higher-complexity ``blips" are more sporadic and much shorter-lived, compared to the regions of high complexity in the helmet streamer boundaries ($HS_1$ and $HS_2$). In addition, there are some intriguing large-scale changes in the boundary during the hour, seen in the panels of Figure \ref{F5} and the associated movie. The third panel of Figure \ref{F5}a and the highlighted region of \ref{F5}b show a transient region of high complexity; inspection of the AIA data shows that this region underwent changes due to a small jet eruption that occurred along the open/closed boundary.

General evaluation of Figure \ref{F5} shows that many regions of the boundary exhibit evidence of near steady-state conditions, without much deviation in dimension value throughout the observed time. A few small areas deviate from this, most notably the jet designated by the boxed region around 300 Mm and from about 7 to 27 minutes, and the aforementioned ``blips" in the pseudostreamer boundary. The boxed area corresponds to the jet just discussed. The most obvious distinction along the boundary is that highlighted by the dashed white lines, showing the helmet streamer and pseudostreamer components of the boundary. This supports the idea put forward by \citet{Aslanyan2022}, that interchange reconnection would occur more readily at a pseudostreamer boundary, smoothing it compared to a helmet streamer boundary, for long time scales. Furthermore, this shows that the reconnection occurs on localized regions and at small scales (in both the temporal and spatial domains). Here, ``small" spatial scales refer not to the diffusion region of the reconnection itself, but to small MHD scales of the reconfiguration after the reconnection. \emph{The short-lived nature and random appearance of the high-dimension ``blips" in the pseudostreamer boundary point strongly to a small-scale stochastic driver for the reconfiguration of the magnetic field.} Transient, localized interchange reconnection is a prime candidate for this mechanism, as we will discuss in the next section.

\section{Discussion and Conclusions}\label{sec:disc}

The main findings of our study can be summarized in the following points:

\begin{enumerate}
    
    \item \textbf{There is a direct correlation between CH boundary complexity and the boundary's bordering closed structure.} We find that there is a distinct and measurable difference in the correlation dimension between the helmet streamer and pseudostreamer boundaries. Not only is the overall boundary of a pseudostreamer generally simpler than that of a helmet streamer, but complex regions that do evolve are quickly smoothed away. In addition, there are more cycles of large-scale complexity changes over the scale of tens of minutes in the helmet streamer boundary. This may mean that the main driver is both time-dependent and spatially-dependent, or that more than one driver is at play. The fundamental implications of these findings present new opportunities for observational estimates of interchange reconnection and major consequences for the stability of open/closed boundaries, which can affect geo-effective space weather events. In subsequent studies, we will expand upon this work to investigate more CHs, longer time spans, and other means of threshold extraction to gain insight into the drivers of CH evolution. 

    \item \textbf{The CDM method showed the most relevant physical scales on the CH boundary are on the order of 5-20 Mm in width; this corresponds to a range of well-studied low coronal structures.} In accordance with \citet{Uritsky2021} and \citet{Kumar2022}, solar coronal plumelets make up highly-structured fine detail in the corona, especially in CHs. These structures are driven by highly dynamic jets -- so-called ``jetlets" \citep{Raouafi2014} -- that occur on scales of 3-5 minutes and 10 Mm. Both the sizes and time spans for these structures correspond well to the lower limit of the evolution scales detected in our present study. This underscores plumelets' importance as sites of interchange reconnection, and as solar wind sources. The upper limit of the relevant spatial scales is consistent with the diameter of supergranules, which are expected to drive much of the corona's structure \citep{Romano2017,Bale2021,Chen2021}. Our results illustrate the ability of the CDM procedure to identify physically important scales. While small and intermediate spatial scales dominate the boundary evolution on relatively short time scales (compared to the lifetime of the average CH), the largest scales may become more important over periods of several days.

    \item \textbf{Since the CDM algorithm is insensitive to the nature of the studied boundary, is represents a powerful new way to quantify irregularities in boundaries of all types.} For CH boundaries, this allows not only the characterization of the entire hole at a given time, but greatly simplifies comparisons over time and between different holes. It is completely unrestricted by data type, meaning that it can be applied without alteration to other wavelength regimes, instruments, and simulation data. There are many potential applications just within solar physics -- CH boundaries, solar flare ribbons, polarity inversion line characterization, and numerical simulation results. 

    \item \textbf{Our analysis supports previous claims that interchange reconnection is the primary driver for CH boundary evolution.} The presented results support the recent predictions of \citet{Aslanyan2022}, and provide strong evidence that the interchange reconnection dynamics could occur at small scales in the corona. One fascinating possibility for these small-scale changes relates to the new findings by Knizhnik et al. (2022; ApJ, in press): the stresses commonly known to create current sheets in the tiny null-point topologies of jets could generate short-lived opposite polarity signatures at the photosphere, driving reconnection and even eruptions. This generation of magnetic field at the photosphere by coronal currents may explain the lack of obvious flux emergence or cancellation in many small jet eruptions -- interchange reconnection is challenging to capture observationally because of its very short time scales, so magnetic field driven by a transient current would likely also be transient. It has also long been postulated that photospheric motions at supergranule boundaries bring magnetic field together and drive reconnection. We can only hint here at the correlation of supergranular flows at photosphere to the evolution reported in this work; a study that aimed to find hard evidence of this correlation would greatly enhance our understanding of the connections between the surface and atmosphere of the Sun \citep[e.g.][]{Attie2015}.

    \item \textbf{This evidence of spatially localized and bursty interchange reconnection would produce magnetic field kink signatures in the upper corona consistent with the switchbacks detected by Parker Solar Probe \citep{Tenerani2020,Zank2020,Bale2021}.} In order to survive and be detected as switchbacks, these would have to survive the plasma $\beta$ transition region in the middle corona \citep{West2022}. Exploring this option is beyond the scope of the current analysis, but is an important research topic for the future.

\end{enumerate}

We have presented and tested the first method to quantify local irregularity of coronal hole boundaries (with the capability of broader applications). The results obtained provide statistical evidence for small-scale transient interchange reconnection operating at CH boundaries, with the pseudostreamer boundary exhibiting evolution on smaller spatial and temporal scales compared to the helmet streamer boundary. These findings lead to new, fundamental questions. Most notably, does the variability reported here occur across all CHs, and how does this affect solar wind creation and release over the lifetime of CH and a solar cycle? We plan to extend this study extensively, to include more CHs, longer observations, and various alternative boundary extraction methods. The application of this powerful new analysis technique to a broader range of boundaries will help characterize reconnection and fast solar wind streams.

\section{Acknowledgements}

This work was part of the ISWAT ``S2-01 Coronal Hole Boundary" Working Team. EIM would like to thank Aleida Higginson, Nicholeen Viall, and Karin Muglach for useful discussions and feedback during this project's development. EIM acknowledges support from the the NASA Heliophysics System Observatory Connect and Living With a Star Programs (grants 80NSSC20K1285 and 80NSSC20K0192). VMU was supported through the Partnership for Heliophysics and Space Environment Research (NASA grant No. 80NSSC21M0180), and acknowledges useful discussions with Peter Grassberger and Olga Uritskaya that helped him understand the value of local fractal metrics. The CDM code was developed by VMU and is available upon request.


\end{document}